\documentclass[aps,prx,twocolumn,english,superscriptaddress]{revtex4-2}

\usepackage{natbib}
\usepackage{amsfonts, amsmath, amssymb, mathrsfs, mathtools, amsthm, bbm, bm}
\usepackage{braket}
\usepackage{float}
\usepackage{afterpage}
\usepackage{color, soul}
\usepackage[version=4]{mhchem}
\usepackage[
    labelfont=bf, 
    font=small,
    justification=justified,
    format=plain
    ]{caption}
\usepackage[font=footnotesize,justification=justified]{subcaption}
\usepackage{comment}
\usepackage{tikz}
\usetikzlibrary{shadows,arrows.meta,positioning,backgrounds,fit,shapes}
\usepackage{quantikz}
\usepackage{booktabs, tabularx, colortbl}
\usepackage[linesnumbered,ruled]{algorithm2e}
\usepackage{quantikz}
\usepackage{blochsphere}
\usepackage{epstopdf}

\AppendGraphicsExtensions{.tga}

\makeatletter
\def\BState{\State\hskip-\ALG@thistlm}
\makeatother

\newcolumntype{Y}{>{\raggedleft\arraybackslash}X}

\theoremstyle{definition}

\newcommand{\angstrom}{\textup{\AA}}

\makeatletter
\renewcommand*\env@matrix[1][*\c@MaxMatrixCols c]{%
  \hskip -\arraycolsep
  \let\@ifnextchar\new@ifnextchar
  \array{#1}}
\makeatother

\usepackage[bookmarks=false]{hyperref}

\begin{document}

\title{\mbox{Towards Compact Wavefunctions from Quantum-Selected Configuration Interaction}}

\author{Tim Weaving}
\email{timothy.weaving.20@ucl.ac.uk}
\affiliation{QMatter, Inc., Office 109, 254 Chapman Rd, Suite 101-B, Newark, Delaware, 19702, USA}
\affiliation{Centre for Computational Science, Department of Chemistry, University College London, WC1H 0AJ, United Kingdom}
\author{Angus Mingare}
\affiliation{Centre for Computational Science, Department of Chemistry, University College London, WC1H 0AJ, United Kingdom}
\author{Alexis Ralli}
\affiliation{QMatter, Inc., Office 109, 254 Chapman Rd, Suite 101-B, Newark, Delaware, 19702, USA}
\affiliation{Centre for Computational Science, Department of Chemistry, University College London, WC1H 0AJ, United Kingdom}
\author{Peter V. Coveney}
\affiliation{Centre for Computational Science, Department of Chemistry, University College London, WC1H 0AJ, United Kingdom}
\affiliation{Advanced Research Computing Centre, University College London, WC1H 0AJ, United Kingdom}
\affiliation{Informatics Institute, University of Amsterdam, Amsterdam, 1098 XH, Netherlands}

\date{\today}

\begin{abstract}
A recent direction in quantum computing for molecular electronic structure sees the use of quantum devices as configuration sampling machines integrated within high-performance computing (HPC) platforms. This appeals to the strengths of both the quantum and classical hardware; where state-sampling is classically hard, the quantum computer can provide computational advantage in the selection of high quality configuration subspaces, while the final molecular energies are evaluated by solving an interaction matrix on HPC and is therefore not corrupted by hardware noise. In this work, we present an algorithm that leverages stochastic Hamiltonian time evolution in Quantum-Selected Configuration Interaction (QSCI), with multireference perturbation theory capturing missed correlations outside the configuration subspace. The approach is validated through a hardware demonstration utilising 42 qubits of an IQM superconducting device to calculate the potential energy curve of the inorganic silane molecule, \ce{SiH4} using a 6-31G atomic orbital basis set, under a stretching of the \ce{Si-H} bond length. We assess the resulting wavefunctions for compactness, a point on which QSCI has previously been criticised. At large separations, where static correlation dominates, we find a configuration space more than $200$ times smaller than that obtained from a conventional SCI selection criterion yields comparable energies. We also compare against the best-in-class Heatbath Configuration Interaction algorithm and observe similar wavefunction compactness at convergence. This result is achieved with a configuration sampling scheme that uses the experimental orbital occupancies of a time-evolved quantum state to predict likely single and double excitations away from existing configurations to bias the subspace expansion procedure. 
\end{abstract}

\maketitle

\section{Introduction}\label{sec1}

Chemistry has been investigated as an application of quantum computing for two decades \cite{aspuru2005simulated} and since 2014 we have seen a series of simulations conducted on hardware architectures based on qubit technologies such as photonics, trapped ions, and superconductors \cite{peruzzo2014variational, Shen2017, OMalleyBabbush2016, Santagati2018a, Kandala2017, Colless2018, hempel2018quantum, kandala2019error, Nam2020, Smart2019, McCaskey2019, Rice2021, arute2020hartree, Gao2021a, Kawashima2021, eddins2022doubling, Yamamoto2022, Kirsopp2022, huang2022variational, lolur2023reference, leyton2022quantum, liang2023napa, Motta2022, OBrien2022, khan2023chemically, zhao2023orbital, guo2023experimental, weaving2023benchmarking, liu2023performing, dimitrov2023pushing, jones2023precision, liang2023spacepulse, weaving2023contextual}. Until the end of 2023, Noisy Intermediate-Scale Quantum (NISQ) simulations of chemistry were dominated by Variational Quantum Algorithms (VQA), which embed a parametrised quantum circuit within a broader classical optimisation scheme. 

\begin{figure*}
\centering
\includegraphics[width=\linewidth]{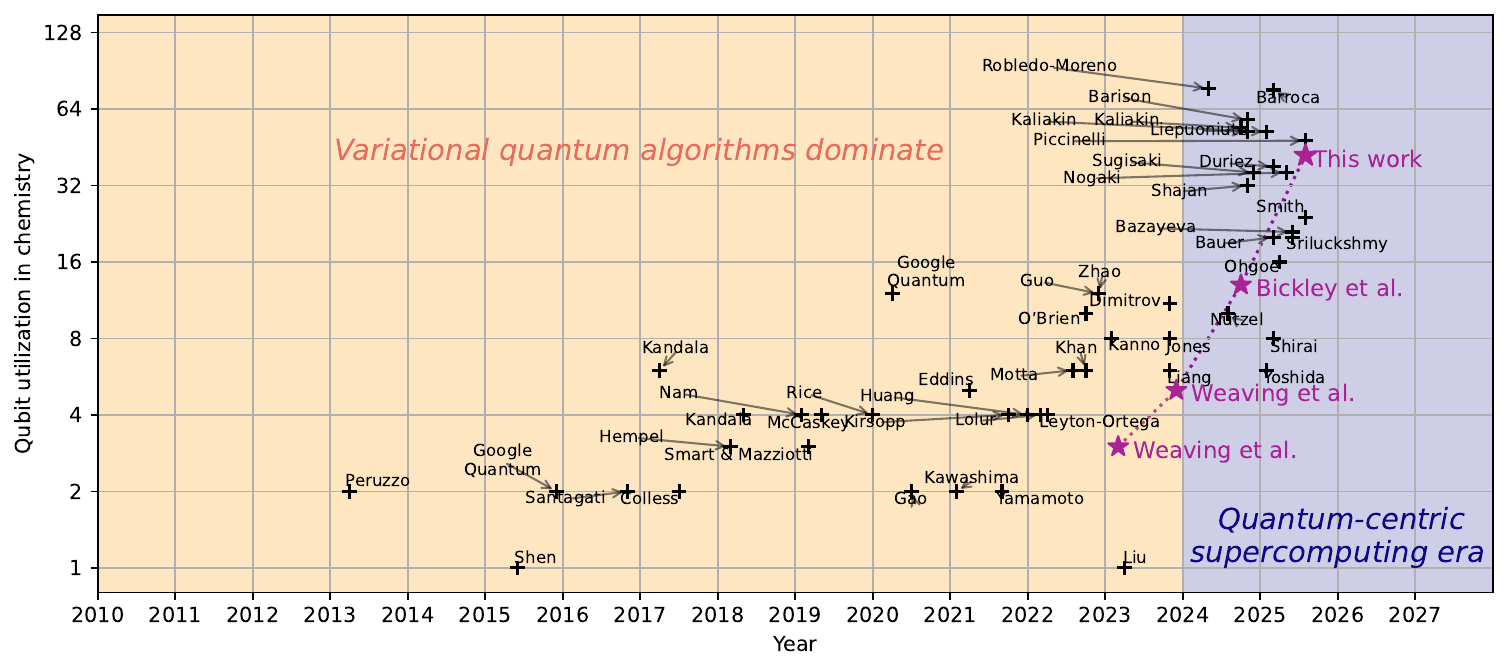}
\caption{Qubit utilisation in chemistry since the beginning of 2013, highlighting the transition from Variational Quantum Algorithms to Quantum-Selected Configuration Interaction techniques after 2024. Dates are taken from the first appearance of the preprint, not the final publication date. Stars indicate work produced by our research group.}
\label{fig:qubit_usage}
\end{figure*}

However, the limited practicality of VQAs meant that NISQ demonstrations had been capped at 12 qubits, as seen in Figure~\ref{fig:qubit_usage}. This can be attributed to several real-world limitations including statistical and hardware noise corrupting the energy estimation procedure, plus the insidious issue of ``barren plateaus'' in the optimisation landscape \cite{mcclean2018barren, holmes2022connecting, cerezo2023does, leone2024practical}. By the end of 2023, it had become increasingly apparent that we needed a paradigm shift in our approach to resolving electronic structure on quantum devices.

A successful framework in conventional electronic structure theory is Selected Configuration Interaction (SCI), where a chosen subset of electronic configurations are interacted and has the ability to capture strong correlation effects, while any missing dynamical (weak) correlation is then treated via multireference perturbation theory. The challenge comes when trying to identify the best electronic states to include in this calculation. However, it has been known for some time that quantum computers excel at state-sampling, while the same is formally hard for classical counterparts \cite{arute2019quantum, zhong2020quantum, wu2021strong, madsen2022quantum}. 

The innovation that is currently propelling us beyond the era of VQAs in chemistry is Quantum-Selected Configuration Interaction (QSCI) \cite{kanno2023quantum, nakagawa2024adapt}, also referred to as Sample-based Quantum Diagonalisation (SQD) \cite{robledo2025chemistry}. Analogous to standard SCI, this consists of sampling configurations from a quantum device, projecting the molecular Hamiltonian into the corresponding subspace and solving classically. Since this hybrid approach can be demanding on the classical compute, it has inspired the term ``quantum-centric supercomputing'' \cite{robledo2025chemistry, alexeev2024quantum} and is positioned to leverage recent trends that see Quantum Processing Units (QPU) co-located with conventional high-performance computing (HPC) platforms~\cite{wintersperger2022qpu, beck2024integrating, bickley2025extending}. 

Unlike VQAs, the inherent hardware noise does not enter the energy calculation step since this is handled classically; it is only the subspace quality that will be influenced by device errors and QSCI is therefore more robust to noise. However, similarly to VQAs, we still have the challenge of constructing effective ansatz circuits. In fact, this is more nuanced in the QSCI setting, since the classically-diagonalised wavefunction need not match that prepared on the quantum device. This observation is central to designing QSCI circuits and should be approached in a different way to VQA ans\"atze.

The earliest QSCI approaches adopted a circuit construction based on the Unitary Cluster Jastrow (UCJ) ansatz \cite{matsuzawa2020jastrow} and localised restrictions to fixed hardware topologies (LUCJ) \cite{motta2023bridging}. The UCJ ansatz approximates the generalised unitary Coupled Cluster wavefunction using $\mathcal{O}(M^2)$ parameters, where $M$ is the number of spatial orbitals. While the LUCJ-QSCI approach has been implemented successfully up to 77-qubits \cite{robledo2025chemistry, kaliakin2025accurate, barison2025quantum, duriez2025computing, barroca2025surface, nogaki2025symmetry, liepuoniute2024quantum, kaliakin2025implicit, bazayeva2025quantum, sriluckshmy2025quantum, shajan2025toward, smith2025quantum, danilov2025enhancing}, it possesses a limitation that casts doubt over the possibility of quantum advantage. The problem is that the LUCJ circuit has variational parameters and thus conceals a VQA -- these can be initialised from cluster amplitudes, but without doing any subsequent quantum optimisation, it seems unlikely to provide benefit over sampling from some classical approximation to the wavefunction. While subsequent work aimed to prove that sampling from LUCJ circuits can be classically-hard in theory \cite{hafid2025hardness}, it has also been demonstrated that the LUCJ circuits seen in the literature to date are amenable to accurate description by tensor networks \cite{rudolph2025simulating}. Aside from LUCJ, several other variational circuit constructions have been proposed for QSCI \cite{yoffe2024qubit, nutzel2025solving, bauer2025efficient, ohgoe2025quantum, shirai2025enhancing}, but have the same limitations. Ideally, our QSCI circuits would be parameter-free to be considered truly ``post-VQA''.

A leading SCI algorithm in conventional quantum chemistry is Heatbath Configuration Interaction (HCI) \cite{holmes2016heat, sharma2017semistochastic, li2018fast}, although it faces a bottleneck in that one must iteratively contract the system Hamiltonian onto a wavefunction approximation to generate an expanded space of connected configurations. While a semistochastic approach was developed to sidestep this issue \cite{sharma2017semistochastic, li2018fast}, one can alternatively replace this Hamiltonian contraction step with Hamiltonian time evolution on a quantum device, an approach that has only recently started to be investigated \cite{sugisaki2024hamiltonian, mikkelsen2024quantum, oumarou2025molecular, Yu2025, piccinelli2025quantum}. This time-evolution variant of QSCI has also been formalised in the framework of Krylov subspaces to facilitate the derivation of theoretical performance guarantees \cite{Yu2025, piccinelli2025quantum}.

Reinholdt et al. criticised QSCI on the basis that repeated sampling can yield an overwhelming duplication of electronic configurations, but this observation arises from the assumption that an ideal circuit should approximate the exact wavefunction as closely as possible. Rather, the most desirable feature of a QSCI circuit is that it be supported over a subset of the dominant configurations in the true wavefunction, but produces those samples close to uniformly. The same authors also argued that, even in the case that electronic configurations are distributed more evenly across the space, QSCI produces less compact expansions compared with competitive heuristics such as HCI \cite{reinholdt2025critical}. 

In this work, we present a configuration sampling heuristic that leverages Hamiltonian time evolution on a superconducting quantum computer to simulate the \ce{SiH4} molecule. The system is described in the split-valence \mbox{6-31G} atomic orbital basis set, consisting of 42 qubits and beyond the minimal basis sets often used in quantum computing demonstrations. While it is recommended to use at least \mbox{6-31G(d,p)} or \mbox{cc-pVDZ} to incorporate polarization functions for this system, it would require 76 qubits in both cases which is too large for the available \texttt{IQM Emerald} device, with 54 qubits. At large \ce{Si-H} separations, where strong, non-dynamical correlation dominates, our methodology produces wavefunctions that are 200 times more compact than a conventional SCI selection criterion that provides a rudimentary upper-bound to the more compact HCI algorithm. While HCI presents a high-bar to exceed, we make headway toward this goal, where further innovations may continue to close this gap until HCI is surpassed by a QSCI-based approach.

In Section~\ref{sec:SCI} we introduce the fundamental concepts of SCI methods, with subsequent multireference perturbation corrections included in Section~\ref{sec:mrpt}. In Section~\ref{sec:evoqsci} we then discuss the motivation behind a Hamiltonian time-evolution approach to QSCI, before describing the stochastic circuit compilation implementation in Section~\ref{sec:qdrift}, later used for our practical demonstration. Specifically, we employ the qDRIFT technique of Campbell \cite{campbell2019random} to construct shallow time evolution circuits, an approach to QSCI developed concurrently by Piccinelli et al. \cite{piccinelli2025quantum}. Finally, we introduce our new configuration sampling algorithm in Section~\ref{sec:sampling_scheme} before outlining the implementation details in Section~\ref{sec:details} and presenting the \ce{SiH4} 6-31G example in Section~\ref{sec:results}.

\section{Methods}

\subsection{Selected Configuration Interaction}\label{sec:SCI}

Configuration Interaction (CI) techniques mix electronic configurations (Slater determinants) to capture correlation effects in molecular systems. Given determinants $\mathcal{D} = \{\ket{\Phi_k}\}_{k=0}^{K-1}$ where $\ket{\Phi_k} = \ket{\bm{b}} \in (\mathbb{C}^2)^{\otimes N}$ for some binary string $\bm{b} \in \mathbb{Z}_2^N$, assumed here to be orthonormal, the configuration subspace projection operator is defined as $P \coloneqq \sum_{k=0}^{K-1} \ket{\Phi_k}\bra{\Phi_k}$. In this subspace, the Hamiltonian $H$ takes the projected form
\begin{equation}
    PHP=\sum_{k,\ell=0}^{K-1} H_{k
    \ell}\;\ket{\Phi_k}\bra{\Phi_\ell},
\end{equation}
specified by a $K\times K$ interaction matrix $\bm{H}$ defined by elements $H_{k\ell} = \bra{\Phi_k}H\ket{\Phi_\ell}$ which may be evaluated efficiently via the Slater-Condon rules \cite{slater1929theory, condon1930theory}. We obtain the eigenvalue equation $\bm{H}\bm{v}_j=\epsilon_j \bm{v}_j$ which produces eigenstates restricted to the configuration subspace $\ket{\Psi_j} = \sum_{k=0}^{K-1} v_{jk} \ket{\Phi_k}$ with energies $\epsilon_j \in \mathbb{R}$, satisfying $PH\ket{\Psi_j} = \epsilon_j \ket{\Psi_j}$. In the more general case that configurations are non-orthonormal, we replace Slater-Condon with the L\"owdin rules \cite{lowdin1955quantum} and must also consider the overlap matrix $\bm{S}$ with $S_{k\ell} = \braket{\Phi_k | \Phi_\ell}$ to solve the generalised eigenvalue equation $\bm{H}\bm{v}_j=\epsilon_j \bm{S} \bm{v}_j$; since we have assumed orthonormality, we have $\bm{S}$ as the identity matrix. In this setting, the development of CI techniques comes down to the way in which we choose $\mathcal{D}$.

For example, if we take $\mathcal{D}$ to be complete -- meaning it contains every possible configuration of $N_\alpha,N_\beta$ spin up/down electrons in $M$ spatial orbitals -- we get Full Configuration Interaction (FCI), the exact solution to the electronic structure problem. The issue is that we have $|\mathcal{D}_{\mathrm{FCI}}| = {M \choose N_\alpha}{M \choose N_\beta}$ and this scales exponentially with increasing numbers of electrons and orbitals. 

Alternatively, if one has knowledge that a subset of orbitals $(M_{\mathrm{act}}, N_{\mathrm{act}})$ -- indicating an active space consisting of $N_{\mathrm{act}} \leq N$ electrons correlated in $M_{\mathrm{act}} \leq M$ orbitals -- is important, then we can limit ourselves to just the configurations connected to this space. Such techniques are referred to as Complete Active Space (CAS) methods and, while the scaling remains exponential in $M_{\mathrm{act}}$, a sufficiently small active space can be chosen to be tractable. 

By limiting ourselves to a combinatorially-restricted subset of configurations, we can simplify the problem to be resolved using polynomially-scaling resources. A common choice is to take the configurations that can be obtained via single or double excitations from the Hartree-Fock reference determinant, producing a subset $\mathcal{D}_{\mathrm{SD}}$ of size $|\mathcal{D}_{\mathrm{SD}}| = {N_\alpha \choose 2}{N_\beta \choose 2}{M-N_\alpha \choose 2}{M-N_\beta \choose 2} + N_\alpha N_\beta (M-N_\alpha) (M-N_\beta) + N_\alpha(M-N_\alpha) + N_\beta(M-N_\beta) + 1$, which can be solved in $\mathcal{O}(M^6)$ using an iterative diagonalisation technique such as Davidson's algorithm \cite{DAVIDSON197587}. 

Some of the most effective methods in advanced electronic structure theory aim to choose the configuration space in a better-informed way. For example, one can treat the full system perturbatively to identify determinants that are ``important'' and then diagonalise the problem in a subspace defined by the dominant contributions in the perturbed wavefunction expansion; this is the approach of Configuration Interaction using a Perturbative Selection made Iteratively (CIPSI) \cite{huron1973iterative, evangelisti1983convergence}. This falls under a class of so-called Selected Configuration Interaction (SCI) methods, which are differentiated by how one assigns importance to each configuration candidate. 

An SCI technique that has achieved considerable success is Heatbath Configuration Interaction (HCI) \mbox{\cite{holmes2016heat, sharma2017semistochastic, li2018fast}}. This works by iteratively expanding the configuration space through solving the Schr\"odinger equation over $\mathcal{D}_i$ (typically initiated with the Hartree-Fock reference, i.e. $\mathcal{D}_0=\{\ket{\Phi_{0}}\}$), yielding eigenstate $\ket{\Psi_i}$, and subsequently contracting the Hamiltonian onto this state, $\ket{\Psi_i} \mapsto H \ket{\Psi_i} = \sum_k \gamma_{ik} \ket{\Phi_k}$. We then select the configurations therein whose coefficients exceed some threshold $\delta>0$ to append to a growing configuration space, namely $\mathcal{D}_{i+1} = \mathcal{D}_{i} \cup \{\ket{\Phi_k}: |\gamma_{ik}|>\delta\}$. This process is iterated until convergence, followed by a perturbation into the full space from the variationally optimised HCI state. As $\delta \rightarrow 0$ HCI approaches the FCI limit since all configurations will then be included.

HCI has been applied to challenging electronic structure problems such as the chromium dimer \ce{Cr2} close to the basis set limit \cite{li2020accurate}, which was later augmented with DMRG data to produce a composite potential energy curve (PEC) for  \ce{Cr2} \cite{larsson2022chromium}. The ability of HCI to capture both strong correlation in the configuration expansion stage, and subsequently capture the missed dynamical (weak) correlation through the final perturbation step, allows the method to produce highly accurate energies and the computational cost can be controlled with the $\delta$ parameter. One problem is the overhead of contracting the Hamiltonian onto the configuration subspace wavefunction $\ket{\Psi_i} \mapsto H \ket{\Psi_i}$ to identify connected determinants for inclusion in the expanded subspace.

This step scales as $\mathcal{O}(L|\mathcal{D}_i|)$ where $L$ is the number of Hamiltonian terms. For smaller systems, most of the molecular integrals are significant and therefore the number of two-electron terms increases as $L=\mathcal{O}(M^4)$. However, as the system size increases substantially, the occurrence of non-negligible orbital overlaps is drastically reduced and consequently the number of relevant two-electron integrals begins to approach $L=\mathcal{O}(M^2)$ \mbox{\cite[p.~401]{strout1995quantitative, helgaker2013molecular}}. Alternatively, this scaling can be reduced via approximate low-rank decomposition methods such as double factorisation \cite{motta2021low, cohn2021quantum} or tensor hypercontraction \cite{lee2021even}. To address the scaling challenge arising from the contraction step, a semistochastic variant of HCI (SHCI) was developed \cite{sharma2017semistochastic, li2018fast}.

\subsection{Multireference Perturbation Theory}\label{sec:mrpt}

Following a CI calculation, as described in the previous Section~\ref{sec:SCI}, we obtain approximate wavefunction(s) $\ket{\Psi_j} = \sum_{\Phi_k \in \mathcal{D}} v_{jk} \ket{\Phi_k}$ with energies $\epsilon_j \in \mathbb{R}$ satisfying $PH\ket{\Psi_j} = \epsilon_j \ket{\Psi_j}$, i.e. eigenstates of the projected Hamiltonian $PHP$. While we can target excited states $j > 0$ through this methodology, in this work we are interested in approximating the ground state $\ket{\Psi_0}$. SCI exactly captures the correlation effects in the restricted configuration subspace spanned by $\mathcal{D}$; however, all remaining correlation lying outside $\mathcal{D}$ is missed. 

One may introduce interactions outside the configuration space through application of perturbation theory. The most convenient formulation for the purposes of SCI is the approach of Epstein-Nesbet \cite{nesbet1955configuration} which takes the model Hamiltonian as
\begin{equation}
    H_0 = \sum_{\Phi_k,\Phi_\ell \in \mathcal{D}} H_{k\ell}\;\ket{\Phi_k}\bra{\Phi_\ell} + \sum_{\Phi_k \notin \mathcal{D}} H_{kk}\;\ket{\Phi_k}\bra{\Phi_k}
\end{equation}
and perturbation operator $V = H - H_0$. We see that the model $H_0$ consists of the full Hamiltonian block inside the configuration subspace and only diagonal elements outside the subspace. By construction, the SCI wavefunctions satisfy $H_0 \ket{\Psi_j} = \epsilon_j \ket{\Psi_j}$, i.e. they are appropriate perturber functions.

Following the standard procedure \cite{song2020multi}, the first-order perturbation correction is $\bra{\Psi_j} V \ket{\Psi_j}=0$ and the second-order correction is
\begin{equation}\label{eq:2nd_corr}
    \epsilon^{(\mathrm{PT2})}_{j} = -\sum_{\Phi_k \notin \mathcal{D}} \frac{|\bra{\Phi_k} V \ket{\Psi_j}|^2}{\bra{\Phi_k} H \ket{\Phi_k}-\epsilon_j},
\end{equation}
yielding $\epsilon^{(\mathrm{SCIPT2})}_j = \epsilon_j + \epsilon^{(\mathrm{PT2})}_{j}$ as the multireference perturbation energy. Computing this correction term has an overhead of $\mathcal{O}(L|\mathcal{D}|)$ and recall from Section~\ref{sec:SCI} this is the same as each iteration of HCI, but here we only incur the cost once, rather than many times as the configuration subspace expands. While the complement space to $\mathcal{D}$ may be exponentially large, one only needs to identify terms with non-zero overlap, which may be achieved through inspection of the contracted state $V \ket{\Psi_j}$.

In the completeness limit $\mathcal{D} \rightarrow \mathcal{D}_{\mathrm{FCI}}$ it is clear the model Hamiltonian $H_0$ will approach the full system Hamiltonian $H$ (in the desired particle sector), so $\epsilon_j(\mathcal{D}) \rightarrow \epsilon^{(\mathrm{FCI})}_j$ and consequently the correction term must decay to zero, $\epsilon^{(\mathrm{PT2})}_{j}(\mathcal{D}) \rightarrow 0$. We can use this fact to motivate an extrapolation scheme by studying the decay of perturbation corrections to estimate the FCI energy. This is investigated later in Section~\ref{sec:results}.

\subsection{Time-Evolved QSCI}\label{sec:evoqsci}

A key computational bottleneck in the Heatbath Configuration Interaction (HCI) algorithm arises when contracting the Hamiltonian onto a reference Slater determinant $\ket{\Phi_k} \mapsto H \ket{\Phi_k}$ to identify connected configurations \cite{holmes2016heat}. This is equivalent to screening all configurations that are related via single and double excitations, of which there are $\mathcal{O}(M^4)$ and therefore scales as $\mathcal{O}(M^4|\mathcal{D}_i)|)$ where $\mathcal{D}_i$ is the configuration subspace at iteration $i$ of the HCI algorithm. To address the scaling challenge, a semistochastic variant of HCI was developed \cite{sharma2017semistochastic, li2018fast}; in our approach, however, this task is delegated to a quantum device.

Since the Hamiltonian is non-unitary, directly implementing its action on a quantum computer would require costly constructions such as block encoding \cite{harrow2009quantum, gilyen2019quantum}, which are impractical for near-term quantum hardware. Instead, we approximate the perturbation by simulating Hamiltonian time evolution using the unitary operator $U = e^{-iHt}$. Decomposing a reference state $\ket{\psi_{\mathrm{ref}}}$ into the eigenbasis $\{{\ket{\Psi_k}}\}_k$ of the Hamiltonian yields
\begin{equation}
\ket{\psi_{\mathrm{ref}}} = \sum_k c_k \ket{\Psi_k} \mapsto \sum_k c_k e^{-i\lambda_k t} \ket{\Psi_k},
\label{eq:expH-perturbation}
\end{equation}
where the expansion coefficients $c_k = \braket{\Psi_k | \psi_{\mathrm{ref}}}$ and $\lambda_k$ is the eigenvalue corresponding to eigenstate $\ket{\Psi_k}$. This operation introduces relative phase shifts between eigenstates, thereby enabling the generation of new configurations that were absent in the initial reference state. We can repeat this process for multiple time steps \mbox{$t = t_0, t_1, \dots, t_K$}, for example with a fixed increment $\tau \in \mathbb{R}$ we take $t_k = k\tau$. The resulting ensemble of sampled configurations is expected to capture the relevant excitations necessary for accurate energy estimation.

This approach has been employed by Sugisaki et al. \cite{sugisaki2024hamiltonian}, Mikkelsen et al. \cite{mikkelsen2024quantum}, Yu et al. \cite{Yu2025} and Piccinelli et al. \cite{piccinelli2025quantum}, the latter two of whom formalise it within the framework of a quantum Krylov subspace method. Classically, a Krylov subspace is constructed from the successive application of the Hamiltonian on a reference state, i.e. $\mathrm{span}\big(\{ \ket{\psi_{\mathrm{ref}}}, H \ket{\psi_{\mathrm{ref}}}, H^2 \ket{\psi_{\mathrm{ref}}}, \dots \}\big)$. In the quantum variant \cite{parrish2019filter, yoshioka2025krylov}, one instead considers repeated application of the unitary operator $U = e^{-iH\tau}$, \mbox{yielding a subspace of the form} 
\begin{equation}    
\mathrm{span}\big(\{\ket{\psi_{\mathrm{ref}}}, U\ket{\psi_{\mathrm{ref}}}, U^2 \ket{\psi_{\mathrm{ref}}}, \dots\}\big).
\end{equation}
Higher powers of $U$ correspond to longer time evolution, $U^k = e^{-iHk\tau}$. For a final time $T$ and a number of steps $K$, we can prepare time-evolved states $\ket{\psi_k} = e^{-iHk\tau}\ket{\psi_{\mathrm{ref}}}$ that propagate by $\tau=T/K$ for $k$ steps on a quantum device and subsequently diagonalise the system in the resulting subspace obtained from measurements of the state. 

\subsection{Stochastic Circuit Compilation}\label{sec:qdrift}

We now turn to the practical implementation of QSCI on quantum hardware. In this work we use the Hartree-Fock determinant \cite{ralli2025bridging} as the initial reference state $\ket{\Phi_0}$ which is trivially prepared with a single layer of Pauli X gates. The challenge then is implementing the unitary time evolution operator $U=e^{-iHt}$. While deterministic product formula approaches such as first-order Trotter decompositions are simple to implement, the circuit depth grows in proportion to the number of terms $L$ in the Hamiltonian \cite{childs2021theory}. As discussed at the end of Section~\ref{sec:SCI}, in the worst case we have $L=\mathcal{O}(M^4)$, although low-rank decomposition methods can reduce this to $L=\mathcal{O}(M^2)$ \cite{motta2021low, cohn2021quantum, lee2021even}. Regardless, for larger systems, this remains prohibitive for NISQ devices. We therefore implement the stochastic compilation technique from Campbell, referred to as qDRIFT \cite{campbell2019random}. A similar approach leveraging qDRIFT was developed concurrently by IBM, as detailed in a recent work by Piccinelli et al. that utilised 48 qubits to simulate an active space of the coronene molecule \cite{piccinelli2025quantum}

In the qDRIFT approach, instead of deterministically simulating each term in the Hamiltonian, the terms are randomly sampled according to the probability distribution defined by their weights. That is, given a Hamiltonian as a weighted sum of Pauli terms, $H = \sum_j h_j \bm{\sigma}_j$, define probabilities $p_j = |h_j|/\lambda$ where $\lambda=\sum_j |h_j|$ is the $\ell_1$ norm of the Hamiltonian. We then randomly sample $N$ terms and append $e^{-i\lambda t \operatorname{sgn}(h_j)\bm{\sigma}_j / N}$ to the decomposition. For a given target precision $\epsilon$ the number of sampled terms is chosen to be $N = \lceil 2\lambda^2t^2/\epsilon\rceil$, hence the circuit depth is effectively decoupled from the number of terms in the Hamiltonian and instead depends on its $\ell_1$ norm, $\lambda$. In the worst case $\lambda$ depends linearly on the number of terms in the Hamiltonian, $L$, however the resulting scaling in $L$ is still favourable compared with Trotter decompositions. Furthermore, due to long-range correlations, this linear scaling is generally not present in quantum chemistry Hamiltonians. For the purposes of QSCI, per time step we can generate many qDRIFT circuit instances, batch submit them to the quantum backend and subsequently collate the retrieved measurements to form our configuration subspace, or perform further postprocessing as discussed in the following section.

\subsection{Configuration Sampling Scheme}\label{sec:sampling_scheme}

A considerable challenge of QSCI is handling hardware noise, but it is difficult to design quantum error mitigation schema since there is a highly nontrivial relationship between the measured bit-strings and the final energy, which involves exact diagonalisation over the configuration subspace spanned by the measurements. By contrast, in algorithms such as VQE, where the energy estimate is obtained via Pauli averaging \cite{peruzzo2014variational}, there is a more direct relationship with the measurement distribution. As a result, it is better motivated to employ mitigation techniques such as Zero-Noise Extrapolation \cite{weaving2023contextual, giurgica2020digital, he2020zero} or Echo Verification \cite{weaving2024accurately}, since the influence of noise on the statistical estimator of interest can be more readily studied and rectified. In QSCI, any technique that aims to fix the measured bit-strings themselves is permitted, for example Readout-Error Mitigation \cite{bravyi2021mitigating, nation2021scalable, van2022model} or error suppression methods such as Dynamical Decoupling \cite{hahn1950spin, viola1998dynamical, facchi2005control, uhrig2007keeping} and Pauli Twirling \cite{geller2013efficient, cai2019constructing}, whereas those that operate by proxy on a statistical estimator are less obviously relevant.

While the above techniques go some way in improving the results obtained from a quantum computer, it is inevitable that hardware noise will still corrupt the device measurements, particularly as we evolve for longer periods of time in our approach to QSCI. The problem here is that, for a given molecular electronic structure problem, it is highly likely the measurements will fall outside the correct particle/spin sector in most instances, even if the circuit itself respects the Hamiltonian symmetries. It is therefore necessary to devise a method of rectifying noisy bit-strings to produce valid electronic configurations that contribute meaningfully to the wavefunction \cite{ivanic2001identification}. 

The \textit{configuration recovery} approach taken by IBM, first presented in the work of Robledo et al. \cite{robledo2025chemistry}, achieves this by probabilistically flipping measured bits towards the classical spin-orbital occupancy distribution of the recovered wavefunction $\ket{\Psi}$ obtained from the previous round of subspace expansion. For a bit-string $\bm{b}$, the probability of flipping bit $b_{\sigma,i}$ -- corresponding to spatial orbital $i \in \{0, \dots, M-1\}$ and spin $\sigma \in \{\alpha, \beta\}$ -- is proportional to the distance $|b_{\sigma,i} - \braket{\Psi| a_{\sigma,i}^\dag a_{\sigma,i} |\Psi}|$ between the current bit value and the average occupancy of that orbital with respect to $\ket{\Psi}$. This recovery procedure mitigates noise-induced deviations in particle number and iteratively refines a statistical prior over configurations, guiding the sampling process toward physically relevant subspaces.

The approach developed for this work does not use the orbital occupancy distribution of intermediate QSCI wavefunctions for configuration recovery, but rather the noisy qubit occupancy distribution measured on the quantum hardware. This distribution is used to predict the most likely single and double excitations above dominant configurations in the QSCI wavefunction from the previous sampling round, generating a new set of configuration that are subsequently screened for inclusion in the expanded configuration subspace. Specifically, for a measurement set $\{\bm{b}_k\}_{k=0}^{N_{\mathrm{shots}}-1}$ we evaluate the probability distribution $P_{\mathrm{occ}}^{(\sigma)}$ of an orbital $\chi_i$ being occupied,
\begin{equation}\label{eq:exp_dist}
    P_{\mathrm{occ}}^{(\sigma)}(\chi_i=1) \propto \frac{1}{N_{\mathrm{shots}}}\sum_{k=0}^{N_{\mathrm{shots}}-1}\braket{\bm{b}_k |a_{\sigma,i}^\dag a_{\sigma,i} | \bm{b}_k}.
\end{equation}

We then iterate over the dominant electronic configurations $\ket{\Phi_k}$ in the previously diagonalised QSCI wavefunction $\ket{\Psi} = \sum_{k} v_k \ket{\Phi_k}$, for example those with a coefficient magnitude exceeding some screening threshold, $|v_k|~>~\epsilon_{\mathrm{screen}}$. Single and double excitations away from $\ket{\Phi_k}$ are then sampled from a probability distribution conditional on the orbital occupancies of $\ket{\Phi_k}$. We select a single excitation $a_p a_q^\dag$ from an occupied index $p$ with probability $P_{\mathrm{occ}}^{(\sigma)}(\chi_p=1|\Phi_{k,p}=1)$ to an unoccupied index $q$ with probability $P_{\mathrm{occ}}^{(\sigma)}(\chi_q=0|\Phi_{k,q}=0)$, so that the overall probably of drawing the index pair $(p,q)$ is given by $P_{\mathrm{occ}}^{(\sigma)}(\chi_p=1|\Phi_{k,p}=1) \cdot P_{\mathrm{occ}}^{(\sigma)}(\chi_q=0|\Phi_{k,q}=0)$. The excited configuration $\ket{\Phi_\ell} = a_p a_q^\dag\ket{\Phi_k}$ is then added to a screening set, and similarly for double excitations $\ket{\Phi_\ell} = a_p a_q a_r^\dag a_s^\dag\ket{\Phi_k}$ from occupied orbital indices $p,q$ to unoccupied indices $r,s$. The probability $P(\Phi_\ell)$ of sampling the configuration $\ket{\Phi_\ell}$ is therefore as follows:
 
\begin{itemize}
    \item Single excitation $\sigma \rightarrow \sigma$, excitation indices $p,q$ selected with probability: $$P_{\mathrm{occ}}^{(\sigma)}(\chi_p=1|\Phi_{k,p}=1) \cdot P_{\mathrm{occ}}^{(\sigma)}(\chi_q=0|\Phi_{k,q}=0)$$
    \item Double excitation $\sigma\sigma \rightarrow \sigma\sigma$ (same-spin), excitation indices $p,q,r,s$ ($p \neq q, r \neq s$) selected with probability:
    \begin{equation*}
    \begin{aligned}
        P_{\mathrm{occ}}^{(\sigma)}(\chi_p=1|\Phi_{k,p}=1) \cdot & P_{\mathrm{occ}}^{(\sigma)}(\chi_q=1|\Phi_{k,q}=1) \cdot \\ P_{\mathrm{occ}}^{(\sigma)}(\chi_r=0|\Phi_{k,r}=0) \cdot & P_{\mathrm{occ}}^{(\sigma)}(\chi_s=0|\Phi_{k,s}=0)
    \end{aligned}
    \end{equation*}
    \item Double excitation $\alpha\beta \rightarrow \alpha\beta$ (opposite-spin), excitation indices $p,q,r,s$ ($p=q, r=s$ allowed) selected with probability:
    \begin{equation*}
    \begin{aligned}
        P_{\mathrm{occ}}^{(\alpha)}(\chi_p=1|\Phi_{k,p}=1) \cdot & P_{\mathrm{occ}}^{(\beta)}(\chi_q=1|\Phi_{k,q}=1) \cdot \\ P_{\mathrm{occ}}^{(\alpha)}(\chi_r=0|\Phi_{k,r}=0) \cdot & P_{\mathrm{occ}}^{(\beta)}(\chi_s=0|\Phi_{k,s}=0)
    \end{aligned}
    \end{equation*}
\end{itemize}

Note only single and double excitations are necessary here as $\braket{\Phi_k| H | \Phi_\ell} = 0$ when configurations differ in more than four spin-orbital positions. Once we have sampled a collection of new configurations $\ket{\Phi_\ell}$ connected to $\ket{\Phi_k}$, they are screened based on the metric
\begin{equation}\label{eq:ranking_metric}
    d(\Phi_\ell) = P(\Phi_l) \cdot |\braket{\Phi_k| H | \Phi_\ell}|
\end{equation}
and we append the configurations with the greatest score to the growing configuration subspace. This configuration sampling scheme is guaranteed to remain within the correct particle sector and can be systematically improved, for example with the screening tolerance, the number of sampling rounds or the number of configurations appended to the subspace at each iteration.

The ranking metric defined in Equation \eqref{eq:ranking_metric} takes inspiration from HCI, where one screens $\ket{\Phi_\ell}$ on the criterion there there exists $\ket{\Phi_k}$ in the current configuration subspace such that $|H_{k\ell}v_k|>\delta$ for some parameter $\delta > 0$ \cite{holmes2016heat, sharma2017semistochastic, li2018fast}. By contrast, in our QSCI approach we effectively replace the wavefunction expansion coefficient $v_k$ with the probability $P(\Phi_\ell)$ derived from a quantum experiment; however, given we only consider excitations away from the configurations $\ket{\Phi_k}$ such that $|v_k|~>~\epsilon_{\mathrm{screen}}$, we do implicitly incorporate the coefficient magnitudes into the selection criteria. The full algorithm is detailed in Algorithm~\ref{sampling_algo} of Appendix~\ref{sec:sampling_scheme_algo}

\section{Implementation Details}\label{sec:details}

\begin{figure}[b]
    \centering
    \includegraphics[width=0.60\linewidth,trim={1cm 1cm 1cm 1cm},clip]{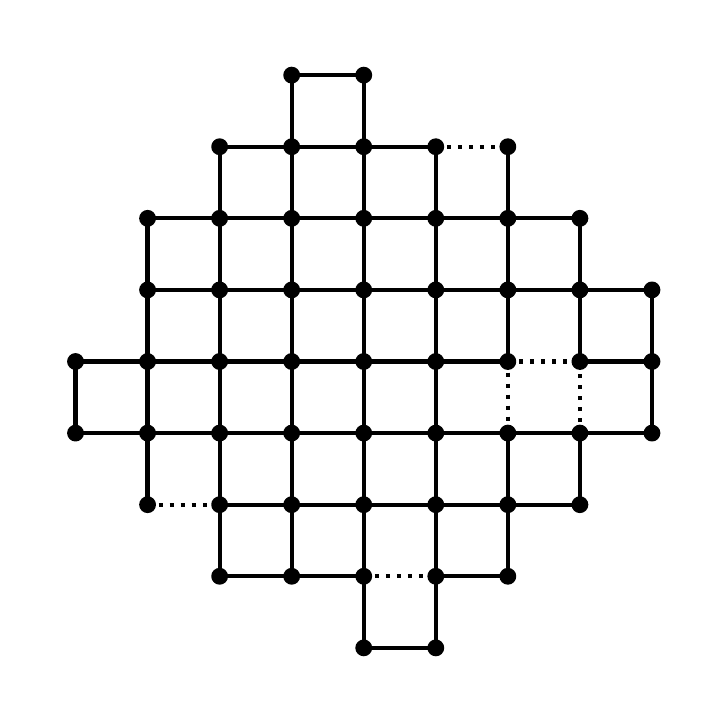}
    \caption{Coupling graph for the \texttt{IQM Emerald} 54-qubit superconducting device. Dotted couplings were not operational at the time of circuit execution.}
    \label{fig:emerald_qpu}
\end{figure}

\begin{figure*}
    \includegraphics[width=\linewidth]{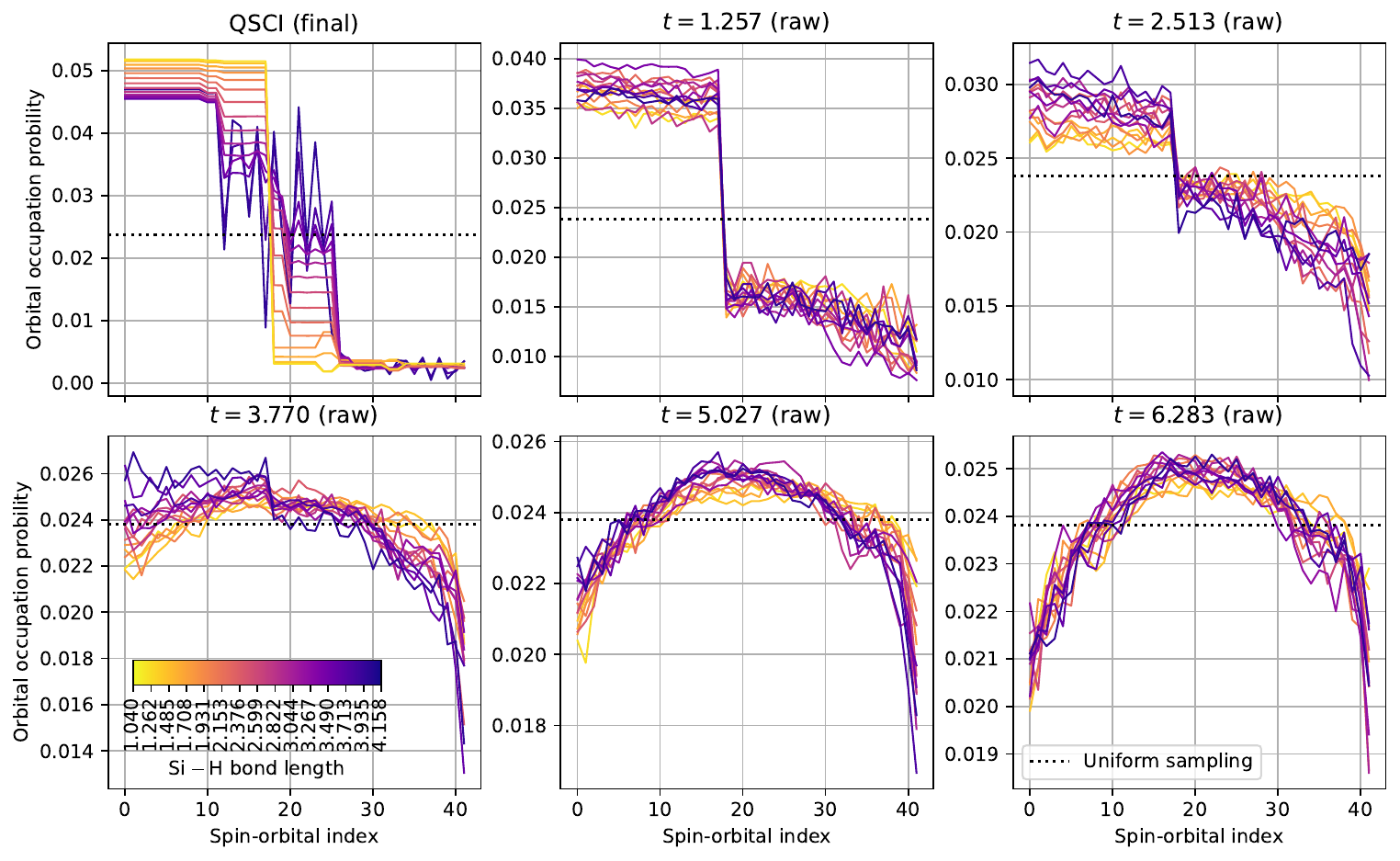}
    \caption{Spin-orbital occupancy distributions for Hamiltonian time evolution applied to the Hartree-Fock reference state on a noisy quantum processor. The time propagation unitary $e^{-iH\tau}$ was applied up to a maximum of five times for increments of $\tau=\frac{2\pi}{5}$ over different \ce{Si-H} bond lengths in the \ce{SiH4} 6-31G system, consisting of 42 qubits. The top left subplot shows the final orbital occupancies obtained by solving the sampled configuration spaces.}
    \label{fig:occ_dists}
\end{figure*}

In the following Section~\ref{sec:results}, we present a practical demonstration of our stochastic Hamiltonian time-evolution approach to Quantum-Selected Configuration Interaction (QSCI) on superconducting quantum hardware, leveraging the configuration sampling scheme of Section~\ref{sec:sampling_scheme}. The device used is the \texttt{IQM Emerald} quantum processing unit (QPU), offering 54 qubits arranged in a square lattice as depicted in Figure~\ref{fig:emerald_qpu}. 

We calculate fifteen points along the potential energy curve (PEC) produced by varying the \ce{Si-H} bond length of \ce{SiH4} 6-31G between $1.040 \angstrom$ and $4.158 \angstrom$. For each point of the PEC we perform five time steps in multiples of $\tau = \frac{2\pi}{5}$ and at every step we generate 50 qDRIFT instances, discussed in Section~\ref{sec:qdrift}, from which $1,024$ measurement shots are extracted. This amounts to $2.560 \times 10^5$ shots per energy estimate and $3.840 \times 10^6$ for the whole experiment. 

After collating measurements from the time evolution stage, we enter the configuration sampling phase; referring to the hyperparameters defined in Algorithm~\ref{sampling_algo}, we set $D_{\mathrm{max}}=5 \times 10^4$ (maximum dimension of configuration subspace), $N_{\mathrm{rounds}}=10$ (number of sampling rounds per measurement set), $N_{\mathrm{samples}}=100$ (number of samples per screened configuration per round), $\epsilon_{\mathrm{screen}} = 10^{-2}$ (configuration screening parameter), $\epsilon_{\mathrm{WF}} = 10^{-5}$ (wavefunction thresholding parameter).

Molecules are built using the \texttt{PySCF} \cite{sun2018pyscf} Python library and Hamiltonian matrix elements are calculated via the Slater-Condon rules \cite{slater1929theory, condon1930theory} using molecular integrals generated by \texttt{libcint} \cite{sun2015libcint}. Interaction matrices are solved using the sparse eigensolver in \texttt{SciPy} \cite{2020SciPy-NMeth}. The qDRIFT Hamiltonian sampling functionality is built on \texttt{symmer} \cite{symmer2022} and converted to time evolution circuits in \texttt{qiskit} \cite{Qiskit}. Quantum jobs are submitted to the \texttt{IQM Resonance} platform for execution on the device. The conventional SCI results are obtained from the selection criterion in \texttt{pyscf.fci.select\_ci}, whereas HCI data were produced with \texttt{PyCI} \cite{richer2024pyci}. 

In both the \texttt{PySCF} SCI and HCI calculations, the variational subspace was expanded using a fixed halving procedure applied to the configuration threshold parameter $\delta$ introduced in Section~\ref{sec:SCI}. Initialised with $\delta = 0.1$, the value of $\delta$ was repeatedly reduced by a factor of two until the subspace reached the maximum permitted dimension. At each threshold level, the configuration space was enlarged through successive Hamiltonian contractions until no further connected configurations satisfying the threshold criterion were identified.

\section{Results}\label{sec:results}

\begin{figure*}
    \centering
    \includegraphics[width=\linewidth]{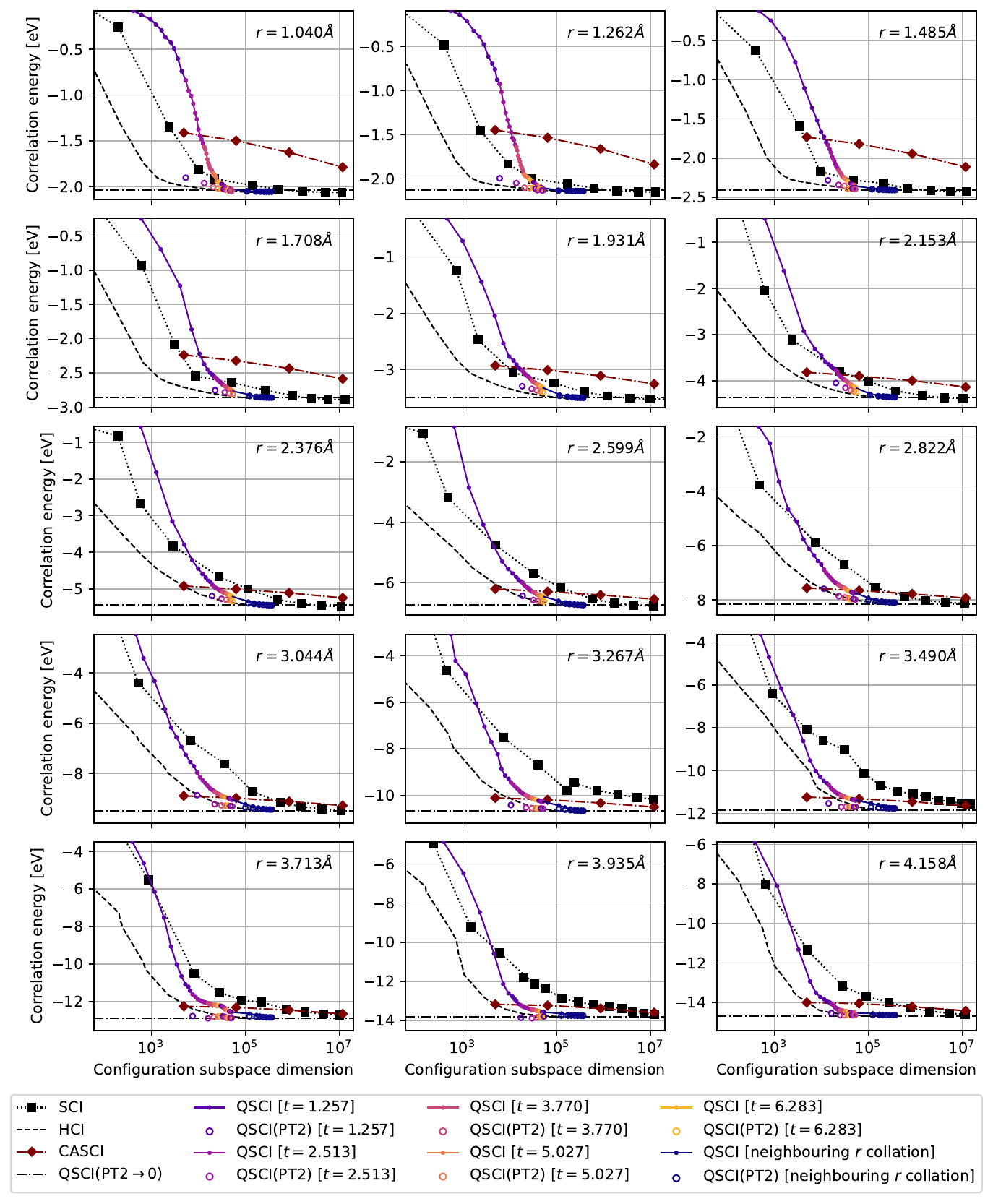}
    \caption{Energy convergence against the size of configuration subspace for our sampling scheme in time-evolved QSCI, compared with the SCI selection criterion in \texttt{PySCF}, the variational stage of HCI as implemented in \texttt{PyCI}, and CASCI calculations over subspaces $(8o,8e), (10o,10e), (12o,12e), (14o,14e)$; the active spaces were selected from the CCSD natural orbitals. The first five colours in the QSCI curve relate to different iterations of the sampling scheme and therefore correspond with the five experimental occupancy distributions in Figure~\ref{fig:occ_dists}, excluding the dark blue points which correspond with collating configurations from neighbouring points in the PEC to further expand the subspace.}
    \label{fig:evoqsci_vs_hci_conv}
\end{figure*}

In this section we present a practical demonstration of our QSCI methodology, as detailed throughout Sections~\ref{sec:SCI} -- \ref{sec:sampling_scheme}. The chosen benchmark system is the inorganic silane compound, consisting of four protons surrounding a single silicon atom in a tetrahedral structure, \ce{SiH4}. We represent the problem in the split-valence \mbox{6-31G} atomic orbital basis set, resulting in a total of 21 spatial orbitals and thus 42 spin-orbitals (qubits) in the molecular Hamiltonian for accommodation on the 54-qubit \mbox{\texttt{IQM Emerald}} superconducting device. The system contains 18 electrons, meaning it is close to the worst-case scenario in terms of total number of valid electronic configurations for a fixed number of spatial orbitals $M=21$, specifically ${M \choose N_\alpha}{M \choose N_\beta} = 8.639 \times 10^{10}$ (the binomial is maximised when $N_\alpha=N_\beta=M/2$). While polarized basis sets such as 6-31G(d,p) or cc-pVDZ would be preferable, they very quickly exceed the size of the chip (76 qubits in that case). The 6-31G(d) basis set would have required 52 qubits and, although within the 54 qubits available, the offline couplings indicated in Figure~\ref{fig:emerald_qpu} would have made circuit construction more challenging.

In Figure~\ref{fig:occ_dists} we observe how the spin-orbital occupancy distribution, given in in Equation \eqref{eq:exp_dist}, changes as we repeatedly apply the time propagator $e^{-iH\tau}$ with $\tau=\frac{2\pi}{5}$ to the Hartree-Fock reference for the \ce{SiH4} 6-31G system on the quantum device. For longer evolution times, with consequently deeper circuits, the distribution approaches uniform sampling (corresponding with fully depolarising noise), although it retains a peaked shape. We observe the maximum of this peak to align with the Fermi level, between the highest occupied and lowest unoccupied molecular orbitals (the HOMO-LUMO gap). While this could be coincidental and simply an artifact of hardware defects, it is opportune with respect to the configuration sampling scheme detailed in Section~\ref{sec:sampling_scheme}, since this will cause excitations nearer the gap to be selected preferentially over those further away. For shorter time evolutions, in this case up to $t = 3.770$ corresponding with three applications of the time propagator, we are able to discern a jump at the HOMO-LUMO gap, rather than the peak observed for later times.

\begin{figure}[bh!]
    \centering
    \begin{subfigure}{\linewidth} 
        \centering
        \includegraphics[width=\linewidth]{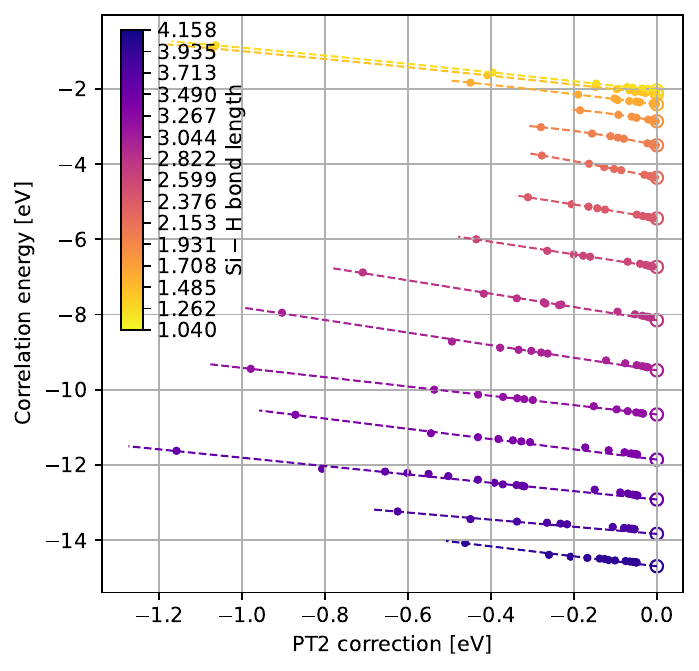}
        \caption{}
        \label{fig:pt2_corrections_extrapolated}
    \end{subfigure}
    \begin{subfigure}{\linewidth} 
        \centering
        \includegraphics[width=\linewidth]{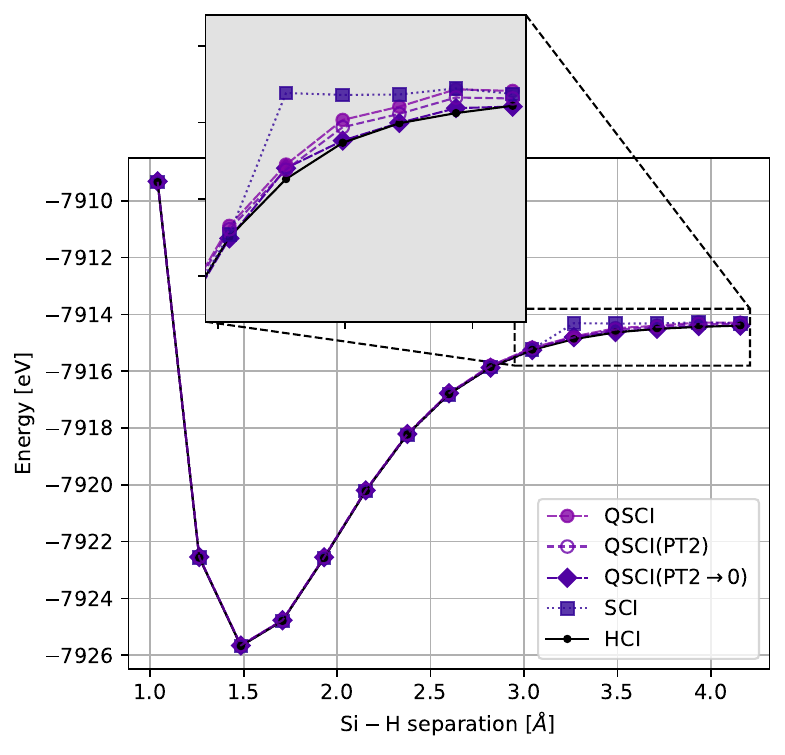}
        \caption{}
        \label{fig:hci_vs_extrapolation}
    \end{subfigure}
    \caption{\textbf{(a)} QSCI correlation energy against second-order perturbation correction. The observed linearity motivates an extrapolation scheme to approximate the FCI energy, where the perturbation correction must decay to zero as the configuration space approaches completeness. \textbf{(b)} Comparing time-evolved QSCI, QSCI(PT2) and the extrapolated QSCI(PT$2\rightarrow0$) curves against SCI/HCI.}
    \label{fig:pt2_extrapolation}
\end{figure}

Using the distributions in Figure~\ref{fig:occ_dists} as the statistical priors that guide the configuration sampling scheme introduced in Section~\ref{sec:sampling_scheme}, we proceed to study the convergence properties of our QSCI algorithm compared against na\"ive SCI, HCI and CASCI. Active spaces for the latter were selected from the CCSD natural orbitals, with sizes $(8o, 8e), (10o, 10e), (12o, 12e), (14o, 14e)$. In Figure~\ref{fig:evoqsci_vs_hci_conv} we find our time-evolved QSCI energies yield similar energies to these other multiconfigurational approaches. Compared against the conventional SCI selection criterion and CASCI results from \texttt{PySCF}, we obtain a considerable reduction in configuration subspace size; this is most pronounced at stretched bond lengths where static correlation dominates, for example from $r=3.267 \angstrom$ onward. 

The largest CASCI space $(14o,14e)$ consists of 
\mbox{$1.178 \times 10^7$} electronic configurations and the largest SCI space solved has size $2.518 \times 10^7$. By contrast, the largest subspace sampled in our QSCI routine consists of just $5.625 \times 10^4$ configurations, $223$ times smaller than the SCI selection criterion in \texttt{PySCF}, representing a considerably more compact expansion of the ground state wavefunction. We also continued expanding the QSCI subspace by collating configurations within an increasing neighbourhood around each point of the PEC, up to a maximum of $3.656 \times 10^5$ (the dark blue points in Figure~\ref{fig:evoqsci_vs_hci_conv}).  However, HCI is the most compact here and remains a formidable target for QSCI-based approaches, as has been true of all preceding literature; the final HCI configuration subspaces are between $1.776\times10^5$ and $4.767\times 10^5$. Despite this, we find that our time-evolved QSCI comes close to HCI in the regime of strong correlation, looking for example at the subplot for $r=4.158 \angstrom$ in Figure~\ref{fig:evoqsci_vs_hci_conv}. This marks substantial progress towards surpassing HCI with a QSCI-based method.

\begin{figure}[b]
    \centering
    \includegraphics[width=\linewidth]{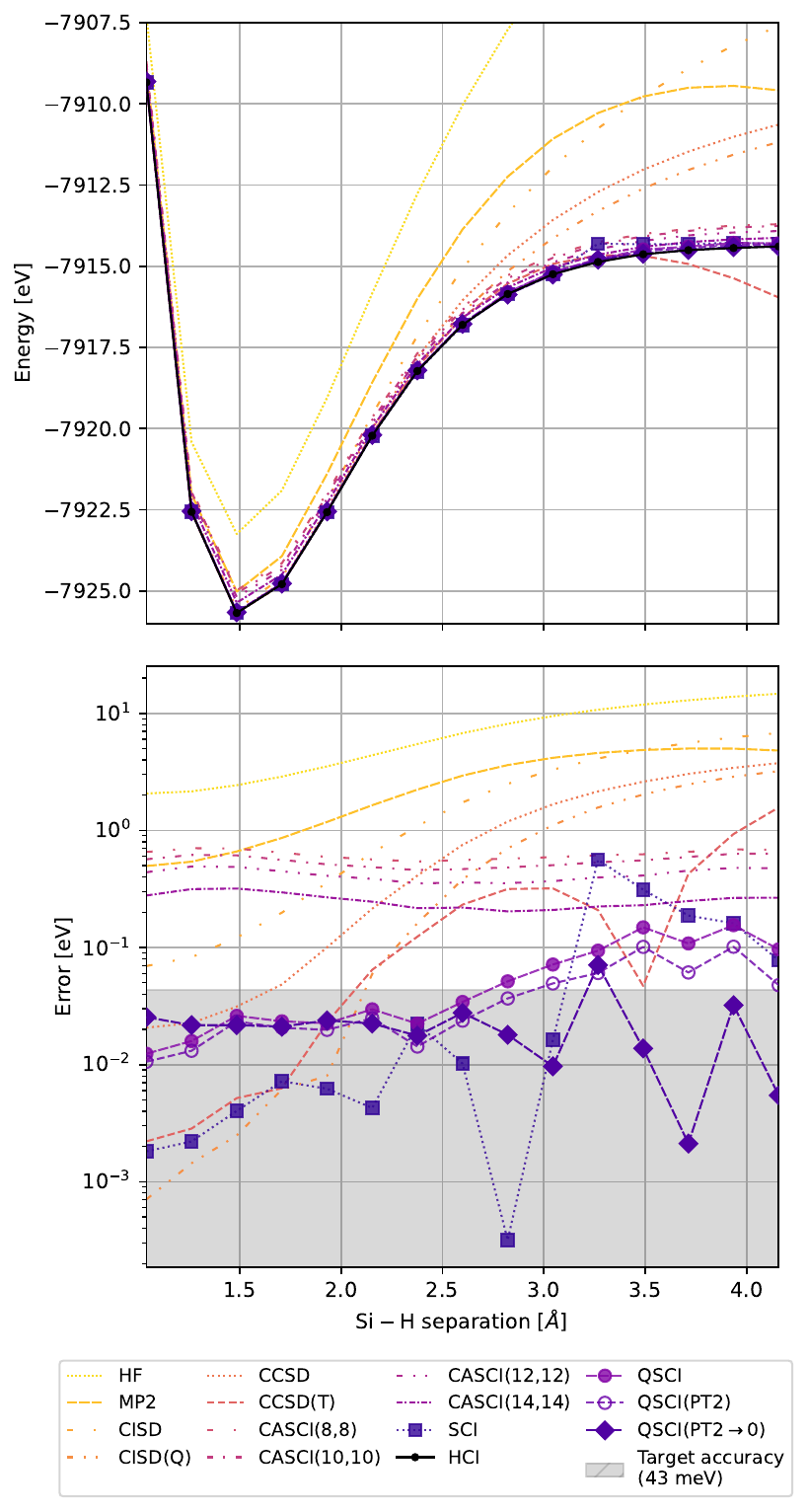}
    \caption{Errors under a stretching of the \ce{Si-H} bond length between $1.040 \angstrom$ and $4.158 \angstrom$, comparing various conventional wavefunction methods against time-evolved QSCI, with and without a multireference perturbation correction, QSCI(PT2), and an extrapolated zero-perturbation energy, QSCI(PT$2\rightarrow0$). Errors are with respect to the HCI energy.}
    \label{fig:final_pec}
\end{figure}

Also plotted in Figure~\ref{fig:evoqsci_vs_hci_conv} are the perturbed QSCI(PT2) energies $\epsilon^{(\mathrm{QSCIPT2})}_0 = \epsilon^{(\mathrm{QSCI})}_0 + \epsilon^{(\mathrm{PT2})}_0$, denoted by hollow points. In Figure~\ref{fig:pt2_corrections_extrapolated} we plot the QSCI correlation energy, defined as the difference $\epsilon^{(\mathrm{QSCI})}_0 - \epsilon_{\mathrm{HF}}$ with the Hartree-Fock energy, against the second-order correction term $\epsilon^{(\mathrm{PT2})}_0$ and find a near-linear relationship. As discussed at the end of Section~\ref{sec:mrpt}, as the configuration space expands towards completeness we must have $\epsilon^{(\mathrm{PT2})}_0 \rightarrow 0$. Therefore, performing regression we may extrapolate to the $\epsilon^{(\mathrm{PT2})}_j=0$ point and take the corresponding value of $\epsilon^{(\mathrm{QSCI})}_0$ as our approximation to $\epsilon_0^{(\mathrm{FCI})}$, denoted QSCI(PT$2\rightarrow0$). In Figure~\ref{fig:hci_vs_extrapolation} we compare the resulting potential energy curve with that obtained from our largest SCI and HCI calculations. The curves are in agreement until a \ce{Si-H} separation of $3.267 \angstrom$, where the \texttt{PySCF} SCI calculation fails to converge, while our extrapolated energies remain smooth throughout and track closely with the HCI curve. 

Finally, in Figure \ref{fig:final_pec} we compare the PEC obtained from time-evolved QSCI, QSCI(PT2) and QSCI(PT$2\rightarrow0$) against the following conventional wavefunction methods: RHF, MP2, CISD, CISD(Q), CCSD, CCSD(T), SCI (\texttt{PySCF} protocol), HCI (\texttt{PyCI}) and CASCI. Errors in the lower subplot are relative to the converged HCI energy.


\section{Conclusion}

In this work we presented a stochastic Hamiltonian time-evolution approach to Quantum-Selected Configuration Interaction (QSCI). Our calculations produced more compact representations of the ground state wavefunction as compared with CASCI and the SCI selection criterion in \texttt{PySCF} by a factor of more than $200$, in terms of the number of configurations included in the subspace. While the criticisms of Reinholdt et al. still stand, namely that the compactness of Heatbath Configuration Interaction (HCI) \cite{holmes2016heat, sharma2017semistochastic, li2018fast} is challenging to improve upon \cite{reinholdt2025critical}, we make considerable progress towards this goal and come close to HCI compactness at stretched bond lengths. We speculate that, with additional modifications to the QSCI implementation, this formidable target can be surpassed.

As discussed in Section~\ref{sec:sampling_scheme}, the new configuration selection criterion developed in this work is related to HCI in that it screens configurations based on Hamiltonian matrix elements, but crucially it incorporates information from a time-evolved quantum state to bias the sampling process with respect to a distribution that is in theory classically inaccessible. While a more extensive benchmark over a suite of challenging molecular systems would be required for thorough validation, the results herein can be interpreted as early evidence for quantum utility in the chemical domain.

\section*{Acknowledgements}
T.W. and A.R. are supported by QMatter, Inc. A.M. acknowledges funding from the Engineering and Physical Sciences Research Council (grant number EP/S021582/1). P.V.C. is grateful for funding from the European Commission for VECMA (800925) and EPSRC for SEAVEA (EP/W007711/1). All authors are grateful to IQM for granting access to their superconducting quantum devices. Particular thanks goes to Eric Mansfield, Stefan Seegerer, Vladimir Kukushkin and Hermanni Heimonen at IQM for providing valuable hardware support throughout the project. Thank you to Peter Reinholdt for bringing \texttt{PyCI} to our attention.

\bibliographystyle{apsrev4-2_mod.bst}
\bibliography{main}

\clearpage
\onecolumngrid
\appendix

\section{Configuration Sampling Scheme}\label{sec:sampling_scheme_algo}

\begin{algorithm}[h!]
    \caption{Symmetry-preserved electronic configuration sampling scheme.}
    \label{sampling_algo}
    
    \SetKwInOut{Input}{Input}
    \SetKwInOut{Output}{Output}

    \Input{Measurements from time-evolution circuits $\mathcal{M} = \{\mathcal{M}_t\}_{t\in\mathcal{T}}$ for times $\mathcal{T}$, where each $\mathcal{M}_t=\{\bm{b}_k\}_{k=0}^{N_{\mathrm{shots}}-1}$ for some number of circuit shots $N_{\mathrm{shots}}\in\mathbb{N}$ and $\bm{b}_k\in\mathbb{Z}_2^{N}$ are binary bit-strings. 
    Hyperparameters: $D_{\mathrm{max}}$ (maximum dimension of configuration subspace), $N_{\mathrm{rounds}}$ (number of sampling rounds per measurement set), $N_{\mathrm{samples}}$ (number of samples per screened configuration per round), $\epsilon_{\mathrm{WF}} > 0$ (wavefunction thresholding parameter), $\epsilon_{\mathrm{screen}} > 0$ (configuration screening parameter) $\delta_{\mathrm{conv}}>0$ (energy convergence tolerance).}
    \Output{Energy $E_0$ and wavefunction $\ket{\Psi_0}$ over an expanded set of electronic configurations $\mathcal{D}$.}
    
    $\mathrm{counter} \leftarrow 0$ {\scriptsize(\textit{counter for sampling rounds})} \;
    $\mathcal{D} \leftarrow \{\Phi_\mathrm{HF}\}$ {\scriptsize(\textit{initial configuration pool})} \;
    $E \leftarrow \braket{\Phi_\mathrm{HF} |H|\Phi_\mathrm{HF} }$ {\scriptsize(\textit{initial energy})} \;
    $\bm{v} \leftarrow (1)$ {\scriptsize(\textit{initial eigenvector, size} $|\bm{v}|=|\mathcal{D}|$)} \;
    $\bm{H} \leftarrow [1]$ {\scriptsize(\textit{initial interaction matrix, size} $|\mathcal{D}|\times|\mathcal{D}|$)} \;
    $\delta \leftarrow 1$ {\scriptsize(\textit{initial convergence parameter})} \;
    \While{$(|\mathcal{D}|<D_{\mathrm{max}}) \wedge (\delta_{\mathrm{conv}} < \delta)$}{
        $n \leftarrow \mathrm{counter} \mod |\mathcal{T}|$ {\scriptsize (\textit{allows looping back to the beginning of the measurement sets again if not yet converged})} \;
        $t \leftarrow \mathcal{T}_n$ \;
        $\mathrm{load}(\mathcal{M}_t)$ {\scriptsize (\textit{read in measurement data for time} $t$)} \;
        $\mathcal{D} \leftarrow \mathcal{D} \cup \{\bm{b}_k\in\mathcal{M}_t:\bm{b}_k \;\text{valid configuration}\}$
        {\scriptsize (\textit{include any bit-strings in the correct particle sector})} \;
        $P_{\mathrm{occ}}^{(\sigma)}(i) \leftarrow \frac{1}{N_{\mathrm{shots}}}\sum_{k=0}^{N_{\mathrm{shots}}-1}\braket{\bm{b}_k |a_{\sigma,i}^\dag a_{\sigma,i} | \bm{b}_k}$ {\scriptsize (\textit{Qubit occupancy numbers for spin} $\sigma\in\{\alpha,\beta\}$, \textit{also normalised for probability})} \;
     
        \For{$n \in \mathrm{range}(N_\mathrm{rounds})$}{
            \For{$\Phi_k\in\mathcal{D}:|v_k|>\epsilon_{\mathrm{screen}}$}{
                \For{$n \in \mathrm{range}(N_\mathrm{samples})$}{
                    $\mathcal{D}_{\mathrm{screen}} \leftarrow \{\}$ {\scriptsize (\textit{initialise empty list for new configurations})} \;
                    \For{$\sigma \in \{\alpha, \beta\}$}{
                        Sample single excitation indices $p,q$ with probability $P(p,q) = P_{\mathrm{occ}}^{(\sigma)}(p | \Phi_{k,p}=1) \cdot P_{\mathrm{occ}}^{(\sigma)}(q | \Phi_{k,q}=0)$
                        \hspace{5cm} {\scriptsize (\textit{i.e. conditional on orbital} $\Phi_{k,p/q}$ \textit{being occupied/unoccupied, respectively})} \;
                        $\mathcal{D}_{\mathrm{screen}} \leftarrow \mathcal{D}_{\mathrm{screen}} \cup \{a_q^\dag a_p\Phi_k\}$ {\scriptsize (\textit{append the new configuration})} \;
                        Sample spin-coupled $\sigma\sigma$ double excitation indices $p,q,r,s$ with probability $P(p,q,r,s) = P_{\mathrm{occ}}^{(\sigma)}(p | \Phi_{k,p}=1) \cdot P_{\mathrm{occ}}^{(\sigma)}(q | \Phi_{k,q}=1) \cdot P_{\mathrm{occ}}^{(\sigma)}(r | \Phi_{k,r}=0) \cdot P_{\mathrm{occ}}^{(\sigma)}(s | \Phi_{k,s}=0)$ \; 
                        $\mathcal{D}_{\mathrm{screen}} \leftarrow \mathcal{D}_{\mathrm{screen}} \cup \{a_r^\dag a_s^\dag a_pa_q\Phi_k\}$ {\scriptsize (\textit{append the new configuration})} \;
                    }
                    Sample spin-paired $\alpha\beta$ double excitation indices $p,q,r,s$ with probability $ P(p,q,r,s) = P_{\mathrm{occ}}^{(\alpha)}(p | \Phi_{k,p}=1) \cdot P_{\mathrm{occ}}^{(\beta)}(q | \Phi_{k,q}=1) \cdot P_{\mathrm{occ}}^{(\alpha)}(r | \Phi_{k,r}=0) \cdot P_{\mathrm{occ}}^{(\beta)}(s | \Phi_{k,s}=0)$ \; 
                    $\mathcal{D}_{\mathrm{screen}} \leftarrow \mathcal{D}_{\mathrm{screen}} \cup \{a_r^\dag a_s^\dag a_pa_q\Phi_k\}$ {\scriptsize (\textit{append the new configuration})} \;
                }
                The above produces $5N_\mathrm{samples}$ configurations for screening, from which we take the top $N_\mathrm{samples}$ determinants $\Phi_\ell$ based on the ranking metric $d:\mathcal{D}_{\mathrm{screen}} \rightarrow \mathbb{R}:d(\Phi_\ell) = P(\Phi_\ell) | \braket{\Phi_\mathrm{k} |H|\Phi_\mathrm{\ell}}|$. 
                Let $\mathcal{D}^\downarrow_{\mathrm{screen}}$ denote the set $\mathcal{D}_{\mathrm{screen}}$ sorted by decreasing $d(\Phi_\ell)$ i.e. for $\Phi_{\ell_i}, \Phi_{\ell_j} \in \mathcal{D}^\downarrow_{\mathrm{screen}}$ with $i<j$ we have $d(\Phi_{\ell_i}) \geq d(\Phi_{\ell_j})$ \;
        
                $\mathcal{D}_{\mathrm{new}} \leftarrow \{\Phi_{\ell_i} \in \mathcal{D}^\downarrow_{\mathrm{screen}}: i=0, \dots,N_{\mathrm{samples}}-1\} \setminus \mathcal{D}$ {\scriptsize (\textit{make sure no duplicate in} $\mathcal{D}$)} \;
            }
            $\bm{H}_{\mathrm{new,new}}, \bm{H}_{\mathrm{new,old}} \leftarrow \begin{bmatrix}
                \braket{\Phi_k | H | \Phi_\ell} & \dots \\
                \vdots & \ddots
            \end{bmatrix}_{\Phi_k,\Phi_\ell \in \mathcal{D}_{\mathrm{new}}}, \begin{bmatrix}
                \braket{\Phi_k | H | \Phi_\ell} & \dots \\
                \vdots & \ddots
            \end{bmatrix}_{\Phi_k \in \mathcal{D}, \Phi_\ell \in \mathcal{D}_{\mathrm{new}}}$ \;
            $\bm{H} \leftarrow \begin{bmatrix}[c|c]
                \bm{H} & \bm{H}_{\mathrm{new, old}} \\ \hline
                \bm{H}_{\mathrm{new, old}}^{\intercal} & \bm{H}_{\mathrm{new, new}}
            \end{bmatrix} $ {\scriptsize (\textit{pad/expand interaction matrix with new configurations})} \;
            $\bm{v} \leftarrow (v_0, \dots v_{|\mathcal{D}|-1}, \underbrace{0, 0, 0, \dots, 0}_{\scriptsize \text{pad} \, |\mathcal{D}_{\mathrm{new}}| \, \text{zeros}})$ \;
            $E_{\mathrm{old}} \leftarrow E$ \;
            $E, \bm{v} \leftarrow \mathrm{eig}(\bm{H}, \bm{v}_\mathrm{init}=\bm{v})$ {\scriptsize (\textit{solve} $\bm{H}\bm{v}=E\bm{v}$, using padded previous $\bm{v}$ \textit{as initial vector in eigensolver})}\;
            $\mathcal{D} \leftarrow \mathcal{D} \cup \mathcal{D}_{\mathrm{new}}$ \;
            $\delta = E_{\mathrm{old}} - E$ {\scriptsize (\textit{noting} $E_{\mathrm{old}} \geq E$ \textit{so} $\delta \geq 0$)} \;
            Filter the configuration set $\mathcal{D}$ by $|v_k| > \epsilon_{\mathrm{WF}}$ \;
        }
        $\mathrm{counter} \leftarrow \mathrm{counter}+1$
    }

\end{algorithm}

\end{document}